\documentclass[12pt,preprint]{aastex} 

\usepackage{graphicx}
\usepackage[twocolumn]{emulateapj5}

\newcommand{\halpha}{H$\alpha$}
\newcommand{\hbeta}{H$\beta$}

\newcommand{\kms}{km~s$^{-1}$}
\newcommand{\flam}{erg~cm$^{-2}$~s$^{-1}$~\AA$^{-1}$}
\newcommand{\sameauthor}{\underbar{\qquad\qquad}.}

\shortauthors{SCHMIDT ET AL.}
\shorttitle{WHITE DWARF/BROWN DWARF BINARY}
\slugcomment{To appear in ApJ (Letters)}
\received{}
\revised{}

\begin{document}

\tolerance 10000

\title{Discovery of a Magnetic White Dwarf/Probable Brown Dwarf Short-Period Binary\altaffilmark{1}}

\altaffiltext{1}{Based in part on observations with the Apache Point
Observatory 3.5~m telescope and the Sloan Digital Sky Survey, which are owned
and operated by the Astrophysical Research Consortium (ARC).}
\author{
Gary D. Schmidt\altaffilmark{2},
Paula Szkody\altaffilmark{3},
Nicole M. Silvestri\altaffilmark{3},
Michael C. Cushing\altaffilmark{2,4},
James Liebert\altaffilmark{2},
and
Paul S. Smith\altaffilmark{2}
}
\vskip 10pt

\altaffiltext{2}{Steward Observatory, The University of Arizona, Tucson, AZ
85721.} \email{gschmidt@as.arizona.edu, mcushing@as.arizona.edu, jliebert@as.arizona.edu, psmith@as.arizona.edu}
\altaffiltext{3}{Department of Astronomy, University of Washington, Box 351580,
Seattle, WA 98195-1580.} \email{szkody@astro.washington.edu, nms@astro.washington.edu}
\altaffiltext{4}{Spitzer Fellow}

\begin{abstract}

The magnetic white dwarf SDSS~J121209.31+013627.7 exhibits a weak, narrow
\halpha\ emission line whose radial velocity and strength are modulated on a
period of $\sim$90 minutes.  Though indicative of irradiation on a nearby
companion, no cool continuum component is evident in the optical spectrum, and
IR photometry limits the absolute magnitude of the companion to $M_J>13.37$.
This is equivalent to an isolated L5 dwarf, with $T_{\rm eff}<1700$~K.
Consideration of possible evolutionary histories suggests that, until $\sim$0.6
Gyr ago, the brown dwarf orbited a $\sim$1.5~$M_\sun$ main seqeunce star with
$P\sim1$~yr, $a\sim1$~AU, thus resembling many of the gaseous superplanets
being found in extrasolar planet searches.  Common envelope evolution when the
massive star left the main sequence reduced the period to only a few hours, and
ensuing angular momentum loss has further degraded the orbit.  The binary is
ripe for additional observations aimed at better studying brown dwarfs and the
effects of irradiation on their structure.

\end{abstract}

\keywords{
stars: low mass, brown dwarfs --- 
stars: individual (SDSS~J121209.31+013627.7) --- 
binaries: close --- 
magnetic fields
}

\section{Introduction}

Detached binary systems consisting of a white dwarf plus a nearby low-mass,
unevolved companion are of importance because they represent the immediate
precursors to cataclysmic variables (CVs) and because their properties allow
inferences into the mechanisms and dependencies of common-envelope (CE)
evolution.  For example, it has been suggested that the lack of magnetic white
dwarf + nonmagnetic main-sequence pairs in current catalogs of detached
binaries could be due, in part, to the role that a magnetic field on the
compact core might play in facilitating the removal of angular momentum
(Lemagie et al. 2004; Liebert et al. 2005).  The classification process itself
is plagued by selection effects, including the larger mean mass (smaller
radius) of magnetic {\em vs.} nonmagnetic white dwarfs (e.g., Liebert et al.
2003), which reduces their detectability against the light of a companion. A
strong magnetic field ($B\gtrsim50$~MG) on a white dwarf also enables the
efficient capture of the wind from a low-mass companion, and the resulting weak
accretion ($\dot M \sim 10^{-13}$ $M_\sun$ yr$^{-1}$) is detectable as
optical/IR cyclotron emission long before the donor's Roche lobe contacts the
stellar surface (Schmidt et al. 2005).  The binary therefore may escape
classification as a detached system, and be included in the Polar class of
magnetic CVs\footnote{It is strictly a ``pre''-Polar.}.  To date, 6 such
binaries have been cataloged from spectroscopic searches like the Sloan Digital
Sky Survey (SDSS).

In this paper we report the discovery and followup observations of a detached
binary with an orbital period of $\sim$90 minutes that contains a magnetic
white dwarf and what appears to be a brown dwarf secondary. The lack of ongoing
accretion and the relative youth of the white dwarf suggest that the components
may have emerged from the CE with an orbital period of only a few hours. The
system thus represents an interesting new wrinkle in the tapestry of binary
star evolution.

\section{Observational Data}

The star SDSS~J121209.31+013627.7 (hereafter SDSS 1212) was reported as a
magnetic white dwarf with an equivalent dipolar magnetic field of $B_d=13$~MG
by Schmidt et al. (2003).  In the 1~hr SDSS spectrum, obtained on 2002 Jan. 8
and reproduced here as Figure 1, very weak emission appears to be present at
\halpha\footnote{And possibly at \ion{Na}{1} D $\lambda$5893, but it is clear
from the near-IR portion of this spectrum that sky-subtraction is not
excellent.}.  This indicates the presence of a nearby companion that is not
apparent as a cool continuum component in the optical spectrum. The emission
line was confirmed through spectroscopy at the APO 3.5~m telescope using the
DIS spectrograph on 2004 Mar. 28 and 2005 May 14, where the regions
$\lambda\lambda4000-5200$ and $\lambda\lambda5950-7600$ were observed in the
two spectrograph channels with a resolution of 2\AA.  Spectropolarimetry was
added on 2005 Apr. 14 and 2005 May 16 with the instrument SPOL (Schmidt et al.
1992) at the Bok 2.3~m telescope, covering the region $4200-8400$~\AA\ at a
resolution of 16\AA.  Though one of the above runs extends as long as 2.2~hr,
no existing data set is ideal for characterizing the behavior of the line
emission due to low time resolution, clouds, or strong moonlight.

The \halpha\ emission and Zeeman-split absorption lines are best displayed in
the APO results from 2005 May 14 shown in Figure 2.  The combined effects of
periodic variations in strength and radial velocity that are apparent in the
right panel cause the emission line to appear doubled when spectra are coadded
over the hour-long sequence.  Separations of the principal triplet components
of \halpha\ and \hbeta\ indicate a mean surface field of $B_s=7$~MG.  A hint of
variation in the shapes of the Zeeman profiles through the series is confirmed
by other data sets, and results in smearing across the higher Balmer lines in
long exposures.

In an effort to detect the companion star in the infrared, imaging was
conducted on 2005 May 20 UT with SpeX, the facility near-IR medium-resolution
spectrograph (Rayner et al. 2003) at the 3.0~m NASA Infrared Telescope Facility
on Mauna Kea.  The spectrograph was configured to an imaging mode using the
guiding camera as the detector, which is equipped with a 512$\times$512 InSb
array, to image SDSS 1212 in the Mauna Kea Observatories Near-Infrared $J$ and
$K$ bands (Tokunaga et al. 2002).  The observations consisted of sequences of
dithered 240~s and 120~s exposures for $J$ and $K$, respectively, using the
UKIRT faint standard FS 132 (Hawarden et al. 2001) for photometric calibration.
The $J$-band measurements were obtained in clear conditions and yielded
$J=17.91\pm0.05$ for 10 integrations.  The $K$-band results exhibit the
systematic effects of rapidly rising humidity that eventually terminated the
observations in fog, and will not be reported here.

\section{Orbital Period}

In all data sets of sufficient length, modulations in the equivalent width (EW)
and radial velocity of the narrow \halpha\ emission line are apparent.
Least-squares sinusoidal fits to the velocities individually yield periods of
$\sim$0.065~d ($\sim$90 minutes), with uncertainties of $\sim$0.01~d.
Unfortunately, at this precision the gaps between observing runs cannot be
bridged without cycle-count ambiguities, so the period cannot be refined at
this time. However, as a consistency check, all data sets were phased onto a
period of 0.065~d and individual phase offsets were applied to register each
set to a common curve.  The results are plotted in the bottom panel of Figure 3
together with a sinusoid of semiamplitude $K_2=320~(\pm20)$~\kms.  The velocity
offset shown, $\gamma=+40$~\kms, is probably not significant.  The shape and
amplitude of the variation are reproducible and consistent with the radial
velocity curve of a low-mass companion orbiting a white dwarf at a rather high
inclination. Indeed, because positive zero-crossing corresponds to inferior
conjunction, disappearance of the emission around this phase (see also Fig.  2)
supports a high inclination and implies that the line emission is confined to
the inner, radiatively-heated hemisphere of the companion.  It might be
expected that the line strength would then peak near superior conjunction
($\varphi=0.5$), but the EW values in the top panel are inconclusive on this
point. The variation in Zeeman structure on a similar timescale as the
emission-line variation that was noted in \S2 suggests that the white dwarf
spin may be synchronized to the orbital period, as in the Polars.  Best
estimates for the pertinent parameters of SDSS 1212 are collected together with
``psf'' magnitudes from the SDSS and our $J$-band photometry in Table 1.

\vspace{.25truein}

\section{The Stellar Components}

By comparing the survey photometry with colors of nonmagnetic DA spectral
models (Bergeron et al. 1995) computed in SDSS filter bands, Schmidt et al.
(2003) estimated a temperature of $T_{\rm eff}=10,000$~K for the white dwarf in
SDSS 1212.  The Zeeman effect tends to increase the EW of (particularly)
\halpha\ and \hbeta, thereby depressing $g$ and to a lesser extent $r$.
However, the magnetic field is also known to alter the continuum opacities, at
least for fields $B\gtrsim100$~MG (Merani et al. 1995). Lacking an adequate
solution for the entire problem for a magnetic white dwarf photosphere, we
adopt the temperature indicated by the nonmagnetic models and quote a liberal
uncertainty of $\pm1,000$~K. The predicted absolute magnitude in SDSS $r$ of
$12.27\pm0.31$ ($\log g = 8$, as indicated by the colors) then implies a
distance modulus $m-M=+5.80\pm0.31$, or $d=145\pm20$~pc, with the error bar
dominated by the uncertainty in temperature.

Absolute $J$-band magnitudes for the same DA white dwarf models range between
$12.29-11.95$ for $T_{\rm eff} = 9,000-11,000$~K, respectively.  Therefore,
with a predicted $J_{\rm wd}=17.90\pm0.14$, the white dwarf alone can account
for the total light of the binary ($J=17.91\pm0.05$) at the coolest end of our
allowed range.  We can set an upper limit for the brightness of the companion
by choosing the hottest permissible white dwarf (largest distance modulus;
$T_{\rm eff}=11,000$~K) and allowing 3$\sigma$ uncertainties on the
photometry, i.e. $J_{\rm min}=17.76$.  With these assumptions, we find that the
companion can contribute at most 21\% of the light at 1.25$\mu$m, or
$J_2>19.44$.  The implied absolute magnitude is $M_{J,2}>13.37$.  By comparison
with the mean characteristics of field L and T dwarfs (Vrba et al. 2004) this
corresponds to a spectral type no earlier than L5, $T_{\rm eff}<1700$K, and
$\log L/L_\sun\le-4.22$.

Of course, the presence of modulated \halpha\ emission signifies the importance
of irradiation by the white dwarf.  We cannot directly utilize the line flux to
estimate a radiating area, but we point out that the blackbody temperature for
reprocessing on the surface of a companion at the implied separation of
0.6~$R_\sun$ is only 1400~K, even assuming an 11,000~K white dwarf, synchronous
rotation, and zero albedo.  Thus, the consideration of radiative heating cannot
yet be used to further constrain the nature of the companion.

The very low inferred luminosity for the companion in SDSS 1212, its large
radial velocity amplitude, the presence of hydrogen at its surface, and the
length of its orbital period argue that the star is a low-mass object near the
cool end of the main sequence, as opposed to, e.g., the eroded core of a
double-degenerate (white dwarf + white dwarf) binary.  The stellar properties
derived from the $J$-band flux are actually near the terminus of
hydrogen-burning stars ($M_2=0.07-0.09~M_\sun$; e.g., Chabrier \& Baraffe
2000), but the fact that the $J$-band luminosity is an upper limit coupled with
the high degree of irradiation prompt us to refer to the companion star as a
brown dwarf.

\section{Nature of the Binary}

We can imagine two possible evolutionary histories for the SDSS 1212 system.
The first takes the binary to be an old Polar in a protracted ($>$3 yr) state
of weak or nonexistent accretion.  EF Eri is a well-known example with
$P=81$~minutes and $B=18$~MG that has been a subject of special interest since
it lapsed into a very low state in 1997.  Though previously thought to contain
an L$4-5$ brown-dwarf companion (Howell \& Ciardi 2001; Harrison et al. 2003),
evidence for IR cyclotron emission by Harrison et al.  (2004) calls this into
question. The white dwarf temperature is similar to that in SDSS 1212 at
$T_{\rm eff}=9,500$~K.

If we make the standard assumption that pre-CVs emerge from the CE at orbital
periods of several hours to days (Taam \& Bodenhemier 1989, 1991; but see Livio
1982 for an alternate route), the evolutionary time to $P\sim90$~minutes is
$>$1 Gyr.  For this age, the radius throughout the L and T sequence is similar
to that of Jupiter (Burrows et al. 1997), and a factor of 2 smaller than the
Roche lobe.  Therefore, if SDSS 1212 is a Polar in a low state, the companion
must be kept artificially inflated during its accretion episodes by radiative
heating.  This picture also requires that the white dwarf exhibit the effects
of accretion-induced heating, since an isolated 0.6~$M_\sun$ white dwarf cools
to $T_{\rm eff}=10,000$~K in $\sim$0.6~Gyr.  Moreover, because accretion only
affects the envelope, the cooling timescale following a lapse of mass transfer
falls in the range weeks to years (Godon \& Sion 2002), so it would seem highly
improbable to have discovered SDSS 1212 in this curious state.

A second, more likely, scenario assumes that the binary is currently and has
always been detached - i.e., that it has never experienced a CV phase.  In this
case the companion was ``born'' as a brown dwarf in an initial orbit with
$P\sim1$~yr, $a\sim1$~AU, i.e., similar to many of the gaseous superplanets
being found in current extrasolar planet surveys.  The orbital period would
have decayed to a few hours during the CE stage, and the binary presumably
has evolved since by the usual gravitational radiation and magnetic braking
angular momentum loss mechanisms.  The post-CE age of the binary is then simply
the $\sim$0.6~GY cooling time of the white dwarf, and the total age of the
brown dwarf is that number added to the few GY lifetime of the main-sequence
parent (for $M_{\rm MS} = 1.5~M_\sun$). This, of course, implies that detached
binaries can emerge from the CE at short orbital periods, a fact already
indicated by population models (Politano 2004) and by the existence of the
2.7~hr period binary nucleus in the planetary nebula Abell 41 (Grauer \& Bond
1983).  While this picture involves fewer phases of evolution, it is no less
interesting, for it offers the possibility of studying the effects of
irradiation on a brown dwarf of known age and history.

\section{Directions for Future Study}

Clearly, $K$-band photometry and spectroscopy are essential.  Unfortunately,
because of its greater distance, detection of the companion in SDSS 1212 may
prove to be more difficult than for EF Eri, which has not yet yielded a clear
result despite the use of the world's largest telescopes.  An improved
\halpha\ radial velocity curve is both desirable and feasible.  An accurate
trigonometric parallax would better constrain the nature of the secondary, but
with the white dwarf dominating even at $J$, accurate model flux distributions
of magnetic degenerate stars are required before the companion's spectrum can
be extracted with confidence. The amplitude of variation of the IR light curve
would allow the effects of irradiation on the brown dwarf to be assessed and
compared with models of the type now under development (e.g., Burrows et al.
2004).  There is also the possibility of a $<$5 minute eclipse of the white
dwarf, a phenomenon that could potentially yield a radius for the companion.
This is amenable to the optical but has not yet been explored through
time-series photometry. Finally, because the temperature of the white dwarf in
SDSS 1212 is very near the range occupied by the ZZ Ceti stars, an accurate
mass for the white dwarf might be available by pulsation analysis.  This
assumes that the magnetic field does not quench the pulsations (Schmidt \&
Grauer 1997; Morsink \& Rezania 2002).  One might even hope to determine the
mass ratio $q=M_2/M_1$ and thus the brown dwarf mass itself by measuring the
time delay in pulsation peaks across the white dwarf's orbit.

\acknowledgements{We are grateful to A. Leistra for assistance with data
analysis.  Support was provided by the NSF through grants AST 03-06080 (G.S.),
AST 02-05875 (P. Szkody and N.S.), and AST 03-07321 (J.L.).  M.C. acknowledges
financial support from the NASA Infrared Telescope Facility and NASA through
the Spitzer Space Telescope Fellowship Program. Funding for the Sloan Digital
Sky Survey has been provided by the Alfred P. Sloan Foundation, the
Participating Institutions, the National Aeronautics and Space Administration,
the National Science Foundation, the U.S.  Department of Energy, the Japanese
Monbukagakusho, and the Max Planck Society.}


\begin{deluxetable}{cl}

\tablecaption{Properties of SDSS~J121209.31+013627.7}

\tablewidth{1.75truein}

\startdata
\hline
\\ [-10pt]
\hline
\\ [-5pt]
$u$ & ~~$18.43\pm0.03$ \\
$g$ & ~~$17.99\pm0.02$ \\
$r$ & ~~$18.07\pm0.02$ \\
$i$ & ~~$18.24\pm0.02$ \\
$z$ & ~~$18.40\pm0.03$ \\
$J$ & ~~$17.91\pm0.05$ \\
$T_{\rm wd}$ & $10,000\pm1,000$~K \\
$B_d$ & ~~~~~~13~MG \\
$P$ & ~~$0.065\pm0.010$~d \\
$K_2$ & ~~~~$\,$$320\pm20$~\kms \\
\enddata

\end{deluxetable}


\begin{figure}
\includegraphics{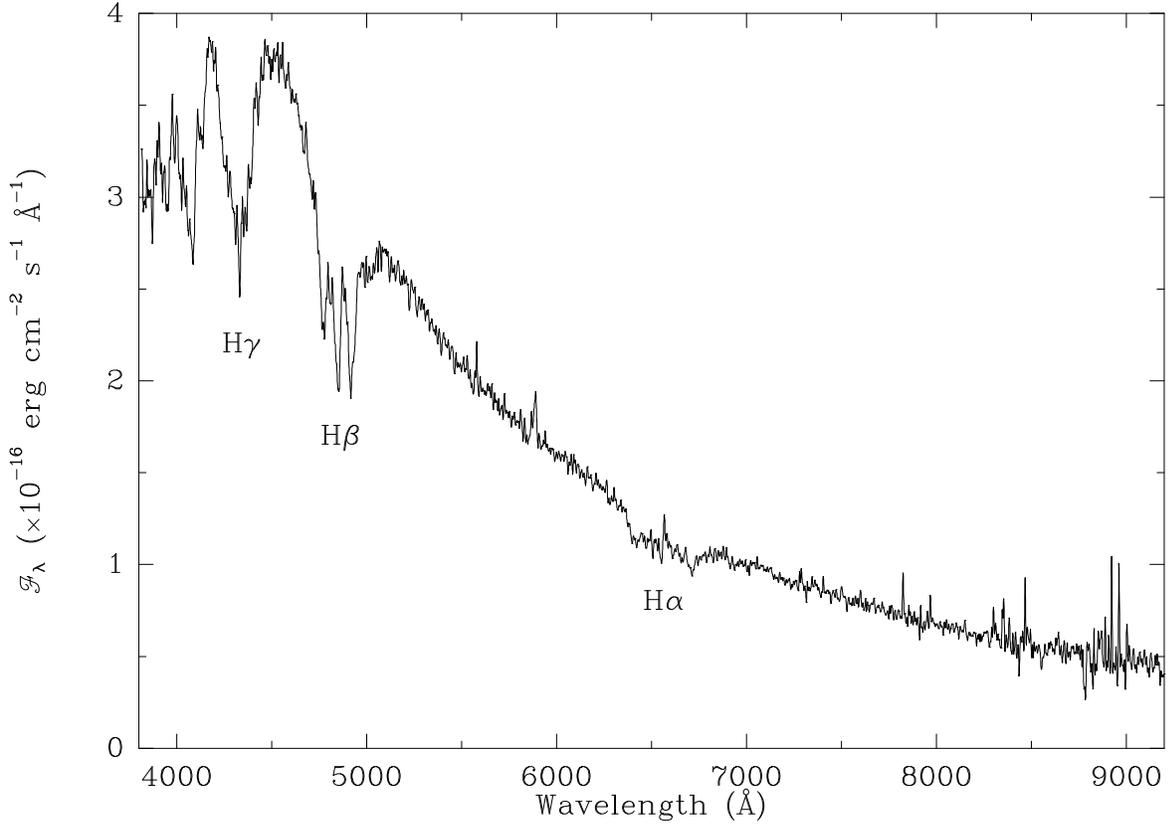}
\vspace{3.truein}

\figcaption{Survey spectrum of SDSS~J121209.31+013627.7 showing Zeeman
splitting in a mean surface field of 7~MG ($B_d=13$~MG).  Note the weak
emission at \halpha\ but lack of evidence for the continuum of a late-type
companion in the red.  Difficulties with sky subtraction are evident beyond
7500\AA\ and may contribute to the appearance of \ion{Na}{1} D weakly in
emission.}
\end{figure}

\clearpage

\begin{figure}
\includegraphics{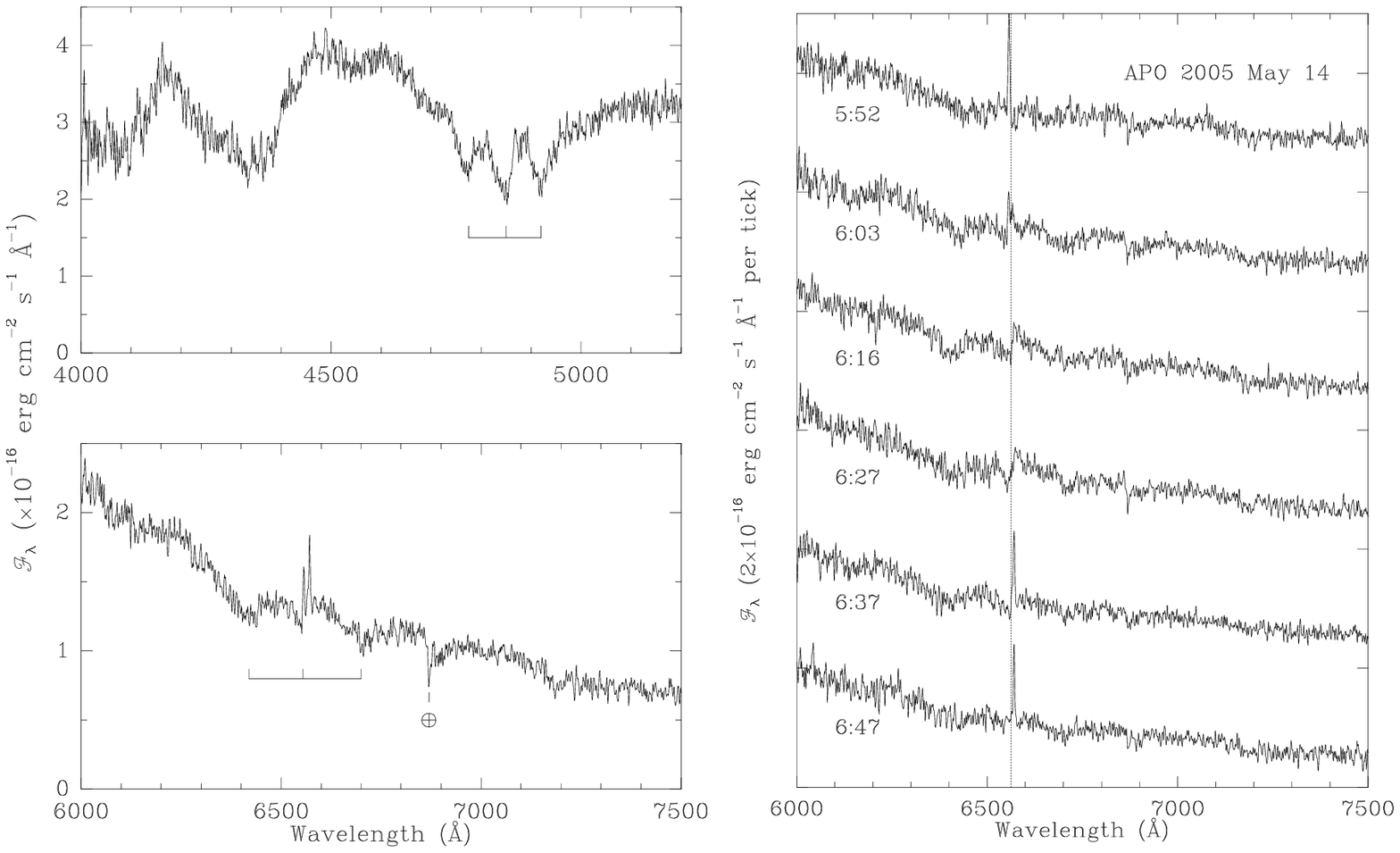}
\vspace{2.truein}

\figcaption{{\em (Right):} One-hour spectroscopic sequence around \halpha\ on
2005 May 14, with the mid-UT of each exposure indicated.  Successive spectra
are displaced by $2\times10^{-16}$~\flam\ and a dotted line at the rest
wavelength of \halpha\ reveals the emission-line velocity shift over the
series.  {\em (Left):} The coadded blue and red spectroscopic channels from the
same data set showing the Zeeman absorption triplets at \halpha\ and \hbeta,
and a doubled appearance of the \halpha\ emission line that arises from the
combined radial velocity and flux modulation of the line.}
\end{figure}

\clearpage

\begin{figure}
\includegraphics{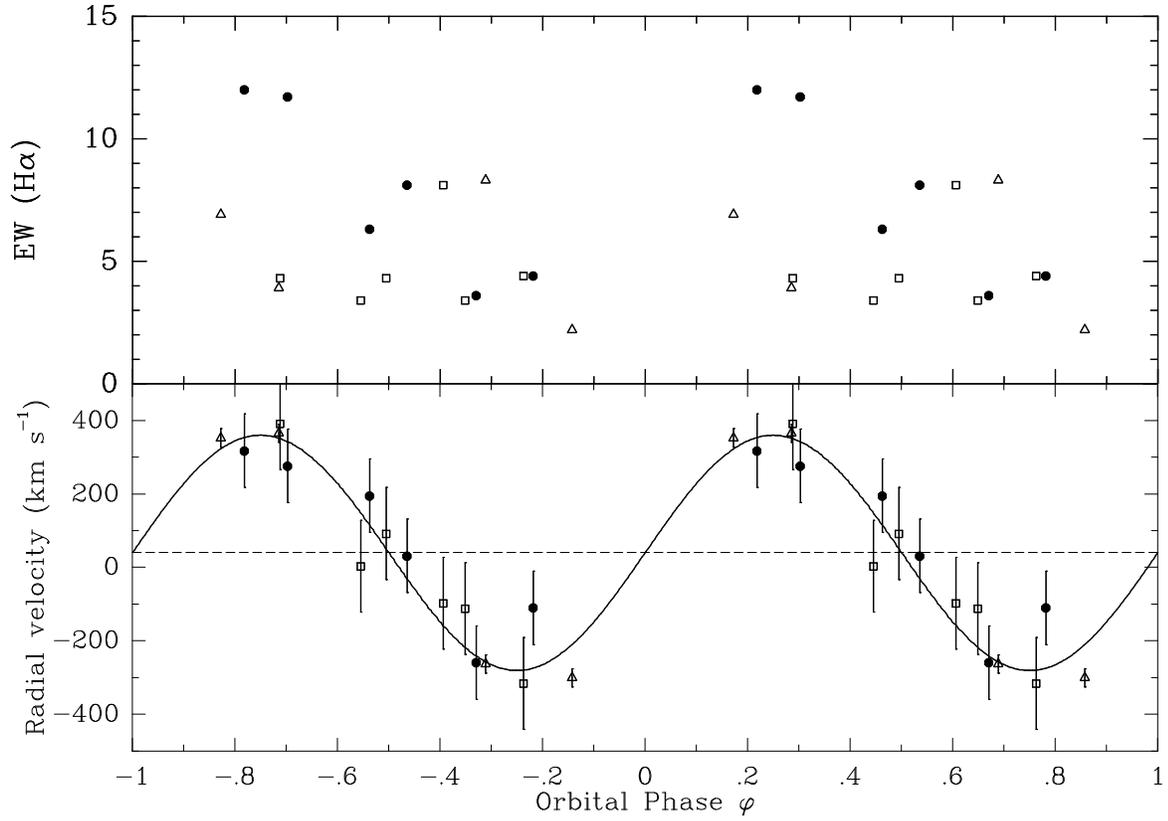}
\vspace{2.5truein}

\figcaption{Radial velocity {\em (bottom)} and equivalent width (EW) {\em
(top)} of \halpha\ for all spectroscopic runs, with individual phase offsets
applied to register the velocity curves to the mean behavior.  The orbital
period is $P=0.065\pm0.01$~d and radial velocity semiamplitude
$K_2=320\pm20$~\kms.  Disappearance of the line around inferior
conjunction ($\varphi=0$) indicates that line emission is confined to the
inner, radiatively-heated hemisphere of the companion, and that the system is
viewed from a high inclination. Symbol key: {\em (filled circles:)} 2004 Mar.
28; {\em (triangles):} 2005 May 14; {\em (squares):} 2005 May 16. The data are
plotted twice for clarity.}
\end{figure}

\end{document}